\documentclass[12pt]{iopart}
\input{psfig}
\begin{document}

\title{Diffraction: A New Approach}

\author{Konstantin Goulianos\dag\footnote{To appear in Proceedings of 
``UK Phenomenology Workshop on Collider Physics, Durham, UK, 19-24 
September 1999," to be published in {\em Journal of 
Physics G: Nuclear and Particle Physics}. This paper is an expanded 
excerpt from hep-ph/9911210, which will apper in Proceedings of
``XXIX International Symposium on Multiparticle Dynamics,
9-13 August 1999, Brown University, Providence, RI 02912, USA",
to be published by {\em World Scientific Publishing Company.}}
}

\address{\dag\ The Rockefeller University, 1230 York Avenue,
New York, NY 10021, USA.\\
e-mail: dino@physics.rockefeller.edu}

\begin{abstract}
A phenomenological model of hard diffraction is presented,
in which the structure of the Pomeron is derived from the
structure of the parent hadron. Predictions for diffractive
deep inelastic scattering are compared with data.
\end{abstract}




%
The inclusive and diffractive deep inelastic scattering (DIS) 
cross sections are proportional to 
the corresponding $F_2$ structure functions of the proton, 

\begin{center}
\begin{tabular}{ll}
Inclusive DIS&$\frac{d^2\sigma}{dxdQ^2}
\propto \frac{F^h_2(x,Q^2)}{x}$\\
Diffractive DIS&
$\frac{d^3\sigma}{d\xi dxdQ^2}
\propto \frac{F^{D(3)}_2(\xi,x,Q^2)}{x}$\\
\end{tabular}
\end{center}
\noindent where the superscripts $h$ and $D(3)$ indicate, respectively, 
a {\em hard} structure function (at scale $Q^2$) and a 3-variable
diffractive structure function (integrated over $t$). 
The latter depends not only 
on the hard scale $Q^2$, but also on the {\em soft} scale,
$\langle M_T\rangle\sim 1$~GeV, which is the relevant scale for the 
formation of the diffractive rapidity gap. 

The only marker of the rapidity gap is the variable $\xi$. We therefore 
postulate that the rapidity gap probability is proportional to the 
{\em soft} parton density at $\xi$ and write the DDIS (diffractive DIS) 
cross section as
$$\frac{d^3\sigma}{d\xi dxdQ^2}\propto \frac{F^h_2(x,Q^2)}{x} \times
\frac{F^s_2(\xi)}{\xi}\otimes \xi{\rm -norm}$$
\noindent where the symbolic notation ``$\otimes \;\xi{\rm -norm}$" is 
used to indicate that the $\xi$ probability is normalized.
Since $x=\beta\xi$, the normalization over all available $\xi$
values involves not only $F_2^s$ but also $F_2^h$, breaking down factorization. 
It is therefore prudent to write the DDIS cross section in terms of $\beta$
instead of $x$, so that the dependence of $F_2^h$ on $\xi$ be shown 
explicitly: 
$$\frac{d^3\sigma}{d\xi d\beta dQ^2}\propto
\frac{1}{\beta}\left[F^h_2(\beta\xi,Q^2)\times
\frac{F^s_2(\xi)}{\xi}\otimes \xi{\rm -norm}\right]
$$
\noindent The term in the brackets represents the DDIS structure function
$F_2^{D(3)}(\xi,\beta,Q^2)$.

In the next step, we seek guidance from the 
scaling behavior of the soft 
single-diffractive (sd) differential cross section~\cite{R,GM},
$$\frac{d\sigma_{sd}}{dM^2}\propto\frac{1}{(M^2)^{1+\epsilon}}
\quad \mbox{(no $s$-dependence!)}$$
\noindent which in terms of $\xi$ takes the form
$$\frac{d\sigma_{sd}}{d\xi}
\propto \underbrace{\frac{1}{s^{2\epsilon}}\;
\frac{1}{\xi^{1+2\epsilon}}}_{\mbox{gap probability}}\times 
(s')^{\epsilon}$$
\noindent where $s'\equiv M^2$ is the s-value of the diffractive sub-system.
Noting that $\xi$ is related to the associated rapidity gap by 
$\Delta Y=\ln\frac{1}{\xi}$, 
and that the integral {\large ${\int}_{s_{\circ}/s}^{1}\;
\frac{1}{s^{2\epsilon}}\;\frac{d\xi}{\xi^{1+2\epsilon}}$} is equal to 
a constant, the above equation may be viewed as representing
the product of the total cross section at the sub-system energy
multiplied by a {\em normalized} rapidity gap probability.
In analogy with this experimentally established behavior,                                                                         
we factorize $F_2^{D(3)}(\xi,\beta,Q^2)$ into $F_2^h(\beta,Q^2)$, 
the sub-energy DIS cross section, times a normalized gap probability:
$$F_2^{D(3)}(\xi,\beta,Q^2)=
P_{gap}(\xi,\beta,Q^2)
\times F^h_2(\beta,Q^2)
\label{F2D3}$$
The gap probability is therefore given by\footnote{In hep-ph/9911210, 
a factor of $\frac1\beta$ was erroneously included in Eq. (13) of the 
gap probability; however, this factor is carried over into the 
normalization factor $N(s,\beta,Q^2)$ 
through Eqs. (14) and (15) and cancels out in the 
final result for $F_2^{D(3)}$ in Eq. (16).}    
$$P_{gap}(\xi,\beta,Q^2)=
F^h_2(\beta\xi,Q^2)\times \frac{F^s_2(\xi)}{\xi}
\times N(s,\beta,Q^2)
\label{GP}$$
The normalization factor, $N(s,\beta,Q^2)$, 
is obtained from the following equation, using $\xi_{min}=Q^2/s$,
$$N^{-1}(s,\beta,Q^2)=\frac{1}{f_q}{\displaystyle\int}_{\xi_{min}}^1
F^h_2(\beta\xi,Q^2)\times \frac{F^s_2(\xi)}{\xi} d\xi$$
\noindent where $f_q$ is the quark fraction of 
the hard structure and is used here 
since only quarks participate in DIS. 

At small $x$~($\le\sim 0.1$), the structure functions $F_2^h$ and $F_2^s$ 
are represented well by the power law expressions~\cite{ZEUSlambda}
$F^h_2(x,Q^2)={A^h}/{x^{\lambda_h(Q^2)}}$ and 
$F^s_2(\xi)={A^s}/{\xi^{\lambda_s}}$. 
Using these forms we obtain
$$N^{-1}(s,\beta,Q^2)=\frac{1}{f_q}\left[
\frac{A^h}{\beta^{\lambda_h}}\;\frac{A^s}{\lambda_h+\lambda_s}\;
\left(\frac{\beta s}{Q^2}\right)^{\lambda_h+\lambda_s}\right]$$
$$F_2^{D(3)}(\xi,\beta,Q^2)=
\frac{1}{\xi^{1+\lambda_h+\lambda_s}}\;
\times f_q(\lambda_h+\lambda_s)\;\left(\frac{Q^2}{\beta s}\right)^
{\lambda_h+\lambda_s}
\times \frac{A^h}{\beta^{\lambda_h}}$$

Since in DDIS $x$ is always smaller than $\xi$, the above form 
of $F_2^{D(3)}$, derived for small $x$, should be valid for all 
$x$ when $\xi$ is small; it should also be valid for all 
$\beta (=x/\xi)$. We therefore expect $F_2^{D(3)}$ to have 
the following $\xi$ and $\beta$ dependence at small $\xi$:
$$F_2^{D(3)}(\xi,\beta,Q^2)|_{\beta,Q^2}\propto \frac{1}{\xi^{1+n}}
\quad\quad n=\lambda_h(Q^2)+\lambda_s$$
$$F_2^{D(3)}(\xi,\beta,Q^2)|_{\xi,Q^2}\propto \frac{1}{\beta^{m}}
\quad\quad m=2\lambda_h(Q^2)+\lambda_s$$
The HERA (non-diffractive) 
DIS measurements~\cite{ZEUSlambda} 
yield $\lambda_s\approx 0.1$, which 
is in agreement with the value of 
$\epsilon=\alpha(0)-1=0.104$~\cite{CMG}, where $\alpha(0)$ is 
the intercept of the Pomeron trajectory at t=0. In the 
$Q^2$ range of 10-50 GeV$^2$, where the DDIS data are concentrated,
these measurements yield
$\lambda_h\approx 0.3$. Using these values we obtain $n=0.4$ and $m=0.7$.
We therefore expect 
$$\mbox{Prediction:}\quad 
F_2^{D(3)}\propto\frac{1}{\xi^{1.4}}\times \frac{1}{\beta^{0.7}}$$

We observe the following features:\\
\vglue 0.5cm
\underline{\bf Factorization}\\
Our prediction exhibits factorization between $\xi$ and $\beta$, 
in agreement with HERA results at small $\xi$.

\vglue 0.5cm
\underline{\bf $\xi$-dependence}\\
In the Regge framework, the $\xi$-dependence of $F_2^{D(3)}$ 
is expected to have the ``Pomeron flux" form of $1/\xi^{1+n}$ 
with $n=2\epsilon=0.2$, independent of $Q^2$. In the $Q^2$ range of 
10-50 GeV$^2$, the HERA experiments find that $n$ is $\approx 0.4$ and has 
a small $Q^2$ dependence, in agreement with our prediction of 
$n=\lambda_h(Q^2)+\lambda_s$. 

\vglue 0.5cm
\underline{\bf $\beta$-dependence}\\
The predicted form  $1/\beta^m$ for $F_2^{D(3)}$ 
is valid in the region of (fixed) small $\xi$ and high $Q^2$,
where the $x$-distribution of $F_2(x,Q^2)$ has the form 
$A^h/x^{\lambda_h(Q^2)}$.
As there are no data points at strictly fixed $\xi$, we have selected 
the following set of five points at $\xi\approx 0.01$ and $Q^2=45$ GeV$^2$
from Ref.~\cite{H1F2D3}:
\begin{center}
\vglue -0.25cm
\begin{tabular}{cccc}
$\beta$&$x$&$\xi=x/\beta$&$\xi\cdot F_2^{D(3)}\pm stat\pm syst$\\
\hline
0.10&0.00133&0.0133&$0.0384\pm 0.0066\pm 0.0030$\\
0.20&0.00237&0.0118&$0.0406\pm 0.0061\pm 0.0026$\\
0.40&0.00421&0.0105&$0.0215\pm 0.0046\pm 0.0016$\\
0.65&0.00750&0.0115&$0.0240\pm 0.0054\pm 0.0026$\\
0.90&0.00750&0.0083&$0.0088\pm 0.0041\pm 0.0005$\\
\end{tabular}
\end{center}
\newpage
Figure~1 shows the above values of 
$\xi \cdot F_2^{D(3)}(\beta)$ versus $\beta$
along with our prediction (solid line).
The following parameters were used in the calculation of $F_2^{D(3)}$:
$\sqrt{s}=280$ GeV, $\xi=0.01$, $Q^2=45$ GeV$^2$, $\lambda_s=0.1$,
$\lambda_h=0.3$, $f_q=0.4$~\cite{CDF_b}, and $A^h=0.2$; the latter was 
evaluated from $F_2(Q^2=50,x=0.00133)=1.46$~\cite{H1F2} 
assuming a $\frac{A^h}{x^{0.3}}$
dependence.
The observed agreement between data and prediction, 
both in shape and normalization,
is considered satisfactory, particularly since no free parameters are used in 
the calculation.
\begin{figure}[h]
\vglue -0.2in
\centerline{\psfig{figure=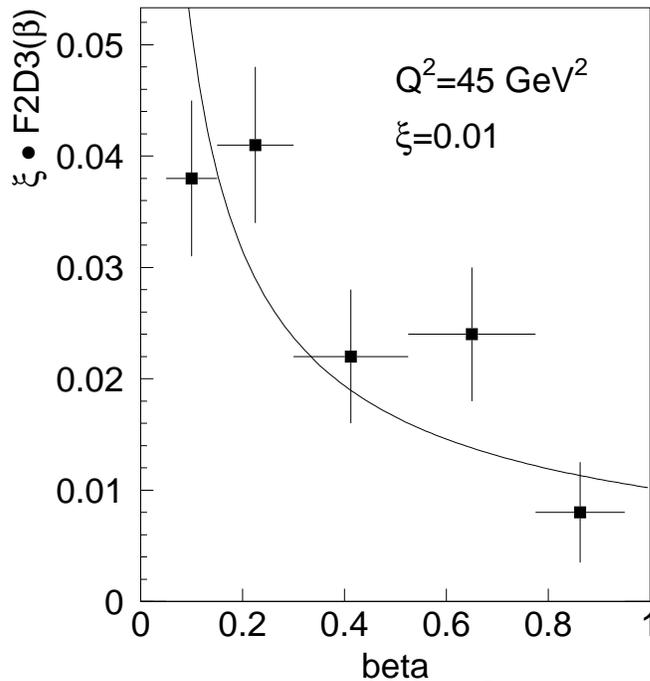,width=4in}}
\vglue -0.2in
\caption{Predicted $\beta$ dependence of 
$\xi\times F_2^{D(3)}(\xi,\beta,Q^2)$ for $\xi=0.01$ and $Q^2=45$ GeV$^2$ 
(solid curve) compared with measured values (points) obtained from 
Ref.~\protect\cite{H1F2D3}}.
\label{fig:H1}
\vglue -0.2in
\end{figure}

\end{document}